# Optimization of graphene dry etching conditions via combined microscopic and spectroscopic analysis

Mariana C. Prado,[1,a)] Deep Jariwala,[2] Tobin J. Marks,[2,3] and Mark C. Hersam[2,3,b)]

[1]Departamento de Física, Universidade Federal de Minas Gerais, Av. Antônio Carlos, 6627, 31270-901 – Belo Horizonte, Brazil.

[2]Department of Materials Science and Engineering, Northwestern University, Evanston, Illinois, 60208, USA.

[3]Department of Chemistry, Northwestern University, Evanston, Illinois, 60208, USA.

**Abstract**

Single-layer graphene structures and devices are commonly defined using reactive ion etching and plasma etching with $O_2$ or Ar as the gaseous etchants. Although optical microscopy and Raman spectroscopy are widely used to determine the appropriate duration of dry etching, additional characterization with atomic force microscopy (AFM) reveals that residual graphene and/or etching byproducts persist beyond the point where the aforementioned methods suggest complete graphene etching. Recognizing that incomplete etching may have deleterious effects on devices and/or downstream processing, AFM characterization is used here to determine optimal etching conditions that eliminate graphene dry etching residues.

[a)] This research was performed while M. Prado was at the Department of Materials Science and Engineering, Northwestern University, Evanston, Illinois, 60208, USA.

[b)] Author to whom correspondence should be addressed. Electronic mail: m-hersam@northwestern.edu

**Manuscript**

Graphene is widely regarded as the prototypical two-dimensional nanomaterial.[1-4] In particular, its unique combination of superlative mechanical[5] and electronic properties[6-8] makes it a promising candidate for electronic applications.[2,6,7,9,10] Since electronic devices require carefully defined geometries, graphene is inevitably subjected to lithographic pattering and etching during fabrication. Furthermore, since single-layer graphene is one-atom thick, any residues from fabrication can significantly influence charge transport and electronic device performance.[11-16] Although several studies have reported dry etching procedures for graphene-based devices,[17-25] this previous work primarily employed optical microscopy and Raman spectroscopy to determine optimal etching conditions. While these methods provide valuable information, they lack sufficient resolution and/or sensitivity to definitively detect residual graphene and/or etching byproducts at the nanoscale.

In this Letter, we supplement optical microscopy and Raman spectroscopy with atomic force microscopy (AFM) to track the evolution of a graphene monolayer following reactive ion etching (RIE) and plasma etching using $O_2$ and Ar. In all cases, AFM reveals that etching residues persist beyond the point where optical microscopy and Raman suggest complete graphene removal. In this manner, we show that AFM allows accurate determination of etching conditions that minimize graphene dry etch residues.

Graphene was mechanically exfoliated onto Si with a 300 nm $SiO_2$ overlayer.[1] Optical microscopy was used to identify the location of single-layer graphene flakes,[26] and Raman spectroscopy enabled subsequent confirmation of the layer multiplicity.[27] To facilitate AFM image acquisition, the samples were pre-cleaned in an Ar/$H_2$ atmosphere at 300°C for 3 hr to remove exfoliation tape residue. Intermittent contact mode AFM images were then acquired in ambient conditions with a Thermomicroscopes CP Research AFM using silicon probes (all-in-one, Budget Sensors) to complete the initial graphene characterization. The graphene samples were subsequently subjected to different etching conditions and characterized with these three techniques after each step. RIE was performed using a Samco RIE-10 NR system with a gas flow rate of 10 sccm, pressure of 26.5 Pa, power of 30 W, and applied DC bias of ~50 V. Plasma cleaning was performed using a Harrick Plasma PDC-001 plasma cleaner at a pressure of 200 mTorr and power of 29.6 W. Although only four samples are reported here, these experiments were repeated over several samples to confirm reproducibility. Additional experimental details are included in the Supplemental Material.[28]

Fig. 1 shows the results for the $O_2$ RIE etched sample. The pristine sample yields a typical graphene monolayer Raman spectrum (Fig. 1a) with two main peaks, G and 2D.[27,29] The 2D peak fits to a single Lorentzian with a full-width-at-half-maximum (FWHM) of 24 cm$^{-1}$, confirming that the flake is single layer.[27,30] No D peak[31] (~1350 cm$^{-1}$) can be detected before the etching process, indicating a low density of defects. The Raman spectra (Fig. 1a) for increasing etching times show a rapid evolution of spectral features, including the appearance of the D peak and reduction of the G and 2D peaks, which imply loss of sp$^2$ character.[31-33] The optical contrast of the monolayer also vanishes with increasing etching time as seen in Figs. 1b, 1d, 1f, 1h. In contrast, the AFM images (Figs. 1c, 1e, 1g) exhibit minimal variation with increasing etching time up to 10 s. In particular, after 10 s of etching, even though there is no discernible Raman signal or optical contrast indicative of single-layer graphene, the AFM image (Fig. 1g) indicates the same topographic contrast as the original sample. Only after 15 s of etching does the graphene appear to be fully removed in the AFM image (Fig. 1i).

While RIE etches quickly due to the directed ion flux, plasma etching requires longer times to remove graphene. Therefore, 1 min was chosen as the $O_2$ plasma treatment increment for each step. Fig. 2a shows the evolution of the Raman spectra, while optical microscopy (Figs. 2b, 2d, 2f, 2h) and AFM (Figs. 2c, 2e, 2g, 2i) images show the changes in optical contrast and topography. After 1 min of plasma exposure, the optical contrast is reduced (Fig. 2d), and the Raman spectrum (Fig. 2a) shows a D peak[33] and no 2D peak. After 2 min of $O_2$ plasma exposure, the flake can no longer be detected with optical microscopy (Fig. 2f) or Raman. However, AFM (Fig. 2g) still shows the appearance of a residual layer. Only after 3 min of plasma etching is the graphene completely removed according to AFM (Fig. 2i).

Since $O_2$ reacts with graphene to form graphene oxide and volatile byproducts, it is a natural choice as an etching gas.[34] In addition, we chose an inert gas, Ar, to compare to the $O_2$ results. Fig. 3a shows the evolution of the Raman spectra following Ar RIE, while optical microscopy (Figs. 3b, 3d, 3f, 3h) and AFM (Figs. 3c, 3e, 3g, 3i) show the changes in optical contrast and topography. After 10 sec, D and G peaks are still detectable. Similarly, AFM detects the remains of the monolayer even though there is no optical contrast. Only after 15 sec do all three techniques detect no monolayer. Although there are differences in the reactivity of Ar with graphene versus that with $O_2$, there is essentially no difference in the RIE etch times under the same conditions. This observation can be attributed to the fact that RIE etching occurs mainly via sputtering and thus the intrinsic of the gases has minimal effect on etch rates.

In contrast to the above results, the effect of chemical reactivity differences between Ar and $O_2$ is evident in the case of plasma etching. In particular, Fig. 4 shows the data for the Ar plasma etched sample. For this case, 2 min steps were employed due to the relatively slow etching rate for the Ar plasma. Fig. 4a shows the evolution of the Raman spectra. The results for optical contrast (Figs. 4b, 4d, 4f, 4 h) and AFM topography (Figs. 4c, 4e, 4g, 4i) remain qualitatively similar to previous samples. For this case, 6 min of etching is sufficient to eliminate the optical contrast (Fig. 4f) and Raman signal. However, as detected by AFM, the remains of the monolayer are only totally removed after 8 min (Fig. 4g). The optical microscopy and AFM images for the 4 min step are included in the Supplemental Material[28] for completeness.

Based on the above observations, we summarize our results for optimal etching times of single-layer graphene in Table 1. Specifically, the etching time that is necessary to see no evidence of the monolayer for each technique is displayed. It is evident that to fully remove the monolayer and any remaining etching residues, the two most widely used techniques to characterize graphene (i.e., optical microscopy and Raman) can be misleading. The characteristic optical contrast and Raman spectrum arising from a single layer of carbon atoms depends on the unique properties of the delocalized electrons in the carbon $sp^2$ honeycomb lattice.[26,27,29] Therefore, if the graphene becomes sufficiently defective or functionalized, it can lose the delocalized $sp^2$ structure, and thus its detectability by optical microscopy or Raman spectroscopy, but it can nevertheless retain the overall structural integrity of a monolayer. Consequently, to rigorously ensure that all graphene and etching residues have been completely removed, nanoscale topographic characterization with AFM is required.

In conclusion, this work shows that optical microscopy and Raman spectroscopy alone are insufficient to determine the complete removal of single-layer graphene and/or etching byproducts following RIE or plasma etching with $O_2$ and Ar. On the other hand, AFM reveals etching residues and thus provides a reliable method for verifying complete etching of single-layer graphene. Due to the single-atom thickness of graphene, we anticipate that full removal of etching residues will impact ongoing efforts to realize more elaborate nanoelectronic devices including heterostructures of disparate two-dimensional nanomaterials.[35,36] Similarly, while this study focuses exclusively on single-layer graphene, this AFM-based approach is also applicable to the etching of thicker graphene samples and other two-dimensional nanomaterials (e.g., transition metal dichalcogenides).

The authors thank Nacional de Grafite for the natural graphite used for exfoliation. This research was supported by the Materials Research Science and Engineering Center (MRSEC) of Northwestern University (NSF DMR-1121262) and by the Office of Naval Research (ONR N00014-11-1-0463). In addition, MCP and MCH acknowledge a CAPES Scholarship (Process Number 5456-11-8) and W. M. Keck Science and Engineering Grant, respectively. This research made use of the NUANCE Center at Northwestern University, which is supported by NSF-NSEC, NSF-MRSEC, Keck Foundation, and the State of Illinois. This research also utilized the NUFAB cleanroom facility at Northwestern University.

[1] K. S. Novoselov, A. K. Geim, S. V. Morozov, D. Jiang, Y. Zhang, S. V. Dubonos, I. V. Grigorieva, and A. A. Firsov, Science **306**, 666 (2004).

[2] F. Schwierz, Nat. Nanotechnol. **5**, 487 (2010).

[3] A. K. Geim and K. S. Novoselov, Nat. Mater. **6**, 183 (2007).

[4] G. W. Flynn, J. Chem. Phys. **135**, 050901 (2011).

[5] C. Lee, X. D. Wei, J. W. Kysar, and J. Hone, Science **321**, 385 (2008).

[6] S. Das Sarma, S. Adam, E. H. Hwang, and E. Rossi, Rev. Mod. Phys. **83**, 407 (2011).

[7] A. H. Castro Neto, F. Guinea, N. M. R. Peres, K. S. Novoselov, and A. K. Geim, Rev. Mod. Phys. **81**, 109 (2009).

[8] D. Zhan, J. X. Yan, L. F. Lai, Z. H. Ni, L. Liu, and Z. X. Shen, Adv. Mater. **24**, 4055 (2012).

[9] A. K. Geim, Science **324**, 1530 (2009).

[10] D. Jariwala, V. K. Sangwan, L. J. Lauhon, T. J. Marks, and M. C. Hersam, Chem. Soc. Rev. **42**, 2824 (2013).

[11] Y.-D. Lim, D.-Y. Lee, T.-Z. Shen, C.-H. Ra, J.-Y. Choi, and W. J. Yoo, ACS Nano **6**, 4410 (2012).

[12] A. Pirkle, J. Chan, A. Venugopal, D. Hinojos, C. W. Magnuson, S. McDonnell, L. Colombo, E. M. Vogel, R. S. Ruoff, and R. M. Wallace, Appl. Phys. Lett. **99**, 122108 (2011).

[13] Y. Dan, Y. Lu, N. J. Kybert, Z. Luo, and A. T. Charlie Johnson, Nano Lett. **9**, 1472 (2009).

[14] M. Ishigami, J. H. Chen, W. G. Cullen, M. S. Fuhrer, and E. D. Williams, Nano Lett. **7**, 1643 (2007).

[15] K. I. Bolotin, K. J. Sikes, Z. Jiang, M. Klima, G. Fudenberg, J. Hone, P. Kim, and H. L. Stormer, Solid State Commun. **146**, 351 (2008).

[16] S. Kumar, N. Peltekis, K. Lee, H.-Y. Kim, and G. Duesberg, Nanoscale Res. Lett. **6**, 390 (2011).

[17] B. J. Lee and G. H. Jeong, Vacuum **87**, 200 (2013).

[18] C. Berger, Z. M. Song, X. B. Li, X. S. Wu, N. Brown, C. Naud, D. Mayou, T. B. Li, J. Hass, A. N. Marchenkov, E. H. Conrad, P. N. First, and W. A. de Heer, Science **312**, 1191 (2006).

[19] K. S. Novoselov, A. K. Geim, S. V. Morozov, D. Jiang, M. I. Katsnelson, I. V. Grigorieva, S. V. Dubonos, and A. A. Firsov, Nature **438**, 197 (2005).

[20] X. S. Wu, Y. K. Hu, M. Ruan, N. K. Madiomanana, J. Hankinson, M. Sprinkle, C. Berger, and W. A. de Heer, App. Phys. Lett. **95,** 223108 (2009).

[21] A. Tzalenchuk, S. Lara-Avila, A. Kalaboukhov, S. Paolillo, M. Syvajarvi, R. Yakimova, O. Kazakova, T. J. B. M. Janssen, V. Fal'ko, and S. Kubatkin, Nat. Nanotechnol. **5**, 186 (2010).

[22] H. M. Wang, Y. H. Wu, Z. H. Ni, and Z. X. Shen, Appl. Phys. Lett.**92,** 053504 (2008).

[23] K. S. Hazra, J. Rafiee, M. A. Rafiee, A. Mathur, S. S. Roy, J. McLauhglin, N. Koratkar, and D. S. Misra, Nanotechnology **22**, 025704 (2011).

[24] N. Peltekis, S. Kumar, N. McEvoy, K. Lee, A. Weidlich, and G. S. Duesberg, Carbon **50**, 395 (2012).

[25] M. S. Choi, S. H. Lee, and W. J. Yoo, J. Appl. Phys. **110**, 073305 (2011).

[26] P. Blake, E. W. Hill, A. H. C. Neto, K. S. Novoselov, D. Jiang, R. Yang, T. J. Booth, and A. K. Geim, Appl. Phys. Lett. **91**, 063124 (2007).

[27] A. C. Ferrari, J. C. Meyer, V. Scardaci, C. Casiraghi, M. Lazzeri, F. Mauri, S. Piscanec, D. Jiang, K. S. Novoselov, S. Roth, and A. K. Geim, Phys. Rev. Lett. **97**, 187401 (2006).

[28] See supplementary material at [URL will be inserted by AIP] for detailed experimental methods and additional Ar plasma etching data.

[29] L. M. Malard, M. A. Pimenta, G. Dresselhaus, and M. S. Dresselhaus, Phys. Rep.-Rev. Sec. Phys. Lett. **473**, 51 (2009).

[30] Z. H. Ni, Y. Y. Wang, T. Yu, and Z. X. Shen, Nano Res. **1**, 273 (2008).

[31] L. G. Cancado, A. Jorio, E. H. M. Ferreira, F. Stavale, C. A. Achete, R. B. Capaz, M. V. O. Moutinho, A. Lombardo, T. S. Kulmala, and A. C. Ferrari, Nano Lett. **11**, 3190 (2011).

[32] M. M. Lucchese, F. Stavale, E. H. Martins Ferreira, C. Vilani, M. V. O. Moutinho, Rodrigo B. Capaz, C. A. Achete, and A. Jorio, Carbon **48**, 1592 (2010).

[33] I. Childres, L. A. Jauregui, J. Tian, and Y. P. Chen, New J. Phys. **13**, 025008 (2011).

[34] L. Liu, S. M. Ryu, M. R. Tomasik, E. Stolyarova, N. Jung, M. S. Hybertsen, M. L. Steigerwald, L. E. Brus, and G. W. Flynn, Nano Lett. **8**, 1965 (2008).

[35] M. S. Choi, G.-H. Lee, Y.-J. Yu, D.-Y. Lee, S. H. Lee, P. Kim, J. Hone, and W. J. Yoo, Nat. Commun. **4**, 1624 (2013).

[36] L. Britnell, R. V. Gorbachev, R. Jalil, B. D. Belle, F. Schedin, A. Mishchenko, T. Georgiou, M. I. Katsnelson, L. Eaves, S. V. Morozov, N. M. R. Peres, J. Leist, A. K. Geim, K. S. Novoselov, and L. A. Ponomarenko, Science **335**, 947 (2012).

TABLE I. Treatment time necessary to see no evidence of the graphene monolayer.

| Technique | RIE | | Plasma | |
|---|---|---|---|---|
| | $O_2$ | Ar | $O_2$ | Ar |
| Optical | 10 s | 10 s | 2 min | 6 min |
| Raman | 10 s | 10 s | 2 min | 6 min |
| AFM | 15 s | 15 s | 3 min | 8 min |

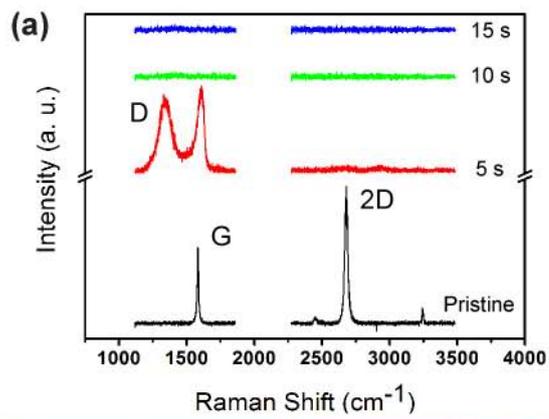
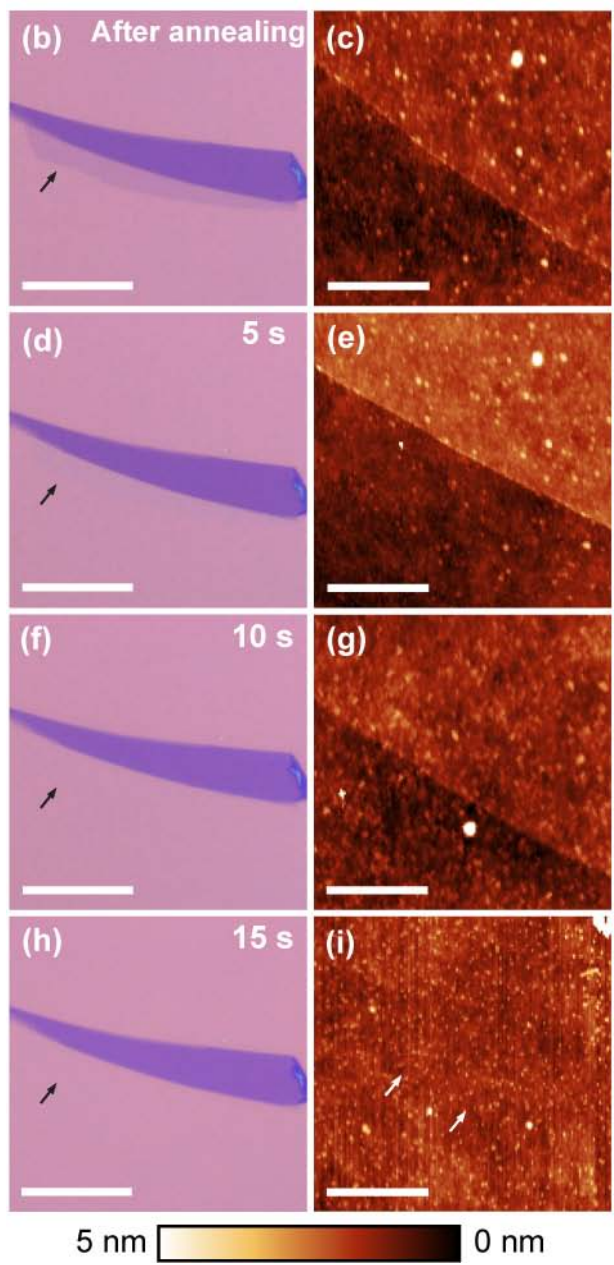

FIG. 1. O$_2$ RIE etching of graphene. (a) Raman data showing the effect of each 5 sec etching pulse. Left column (b, d, f, h) are optical microscopy images of the graphene flake following successive 5 sec etching pulses (scale bar: 20 µm). Right column (c, e, g, i) are AFM images of the graphene flake, taken in the region indicated by the black arrow in the corresponding optical image (scale bar: 1 µm). In (i), the white arrows indicate the original location of the monolayer edge.

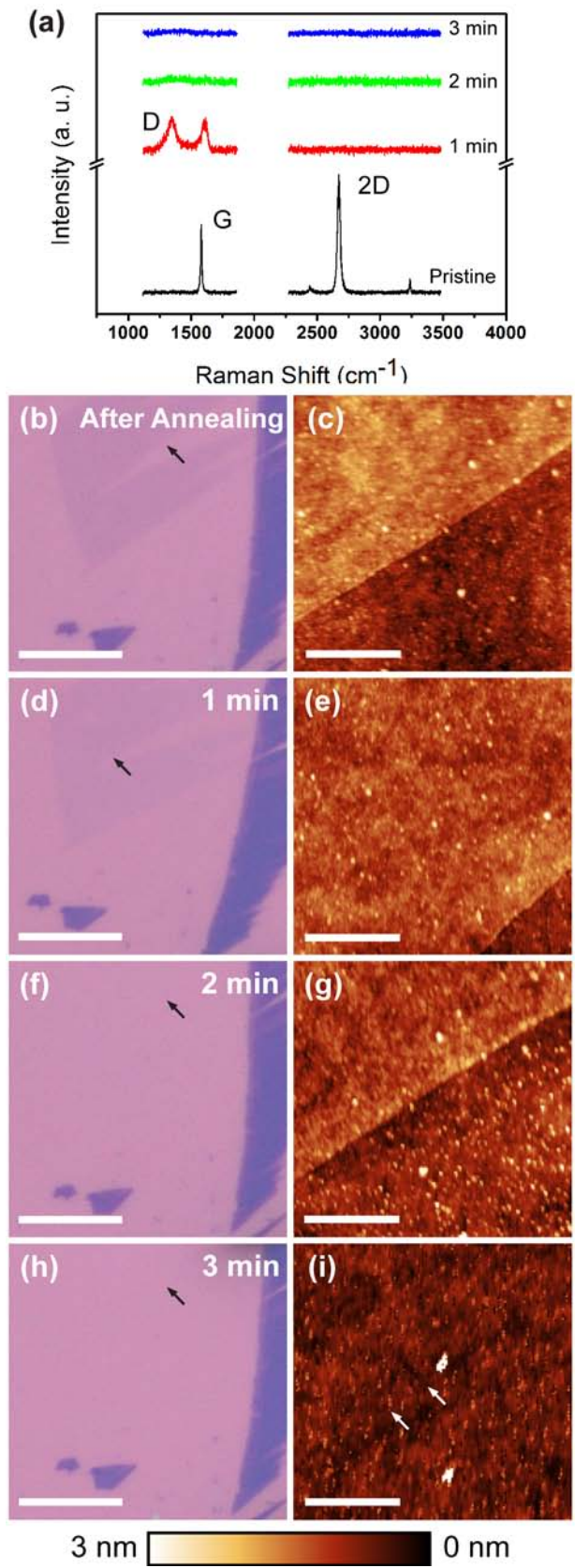

FIG. 2. O$_2$ plasma etching of graphene. (a) Raman data showing the effect of each 1 min etching pulse. Left column (b, d, f, h) are optical microscopy images of the graphene flake

following successive 1 min etching pulses (scale bar: 10 µm). Right column (c, e, g, i) are AFM images of the graphene flake, taken in the region indicated by the black arrow in the corresponding optical image (scale bar: 1 µm). In (i), the white arrows indicate the original location of the monolayer edge.

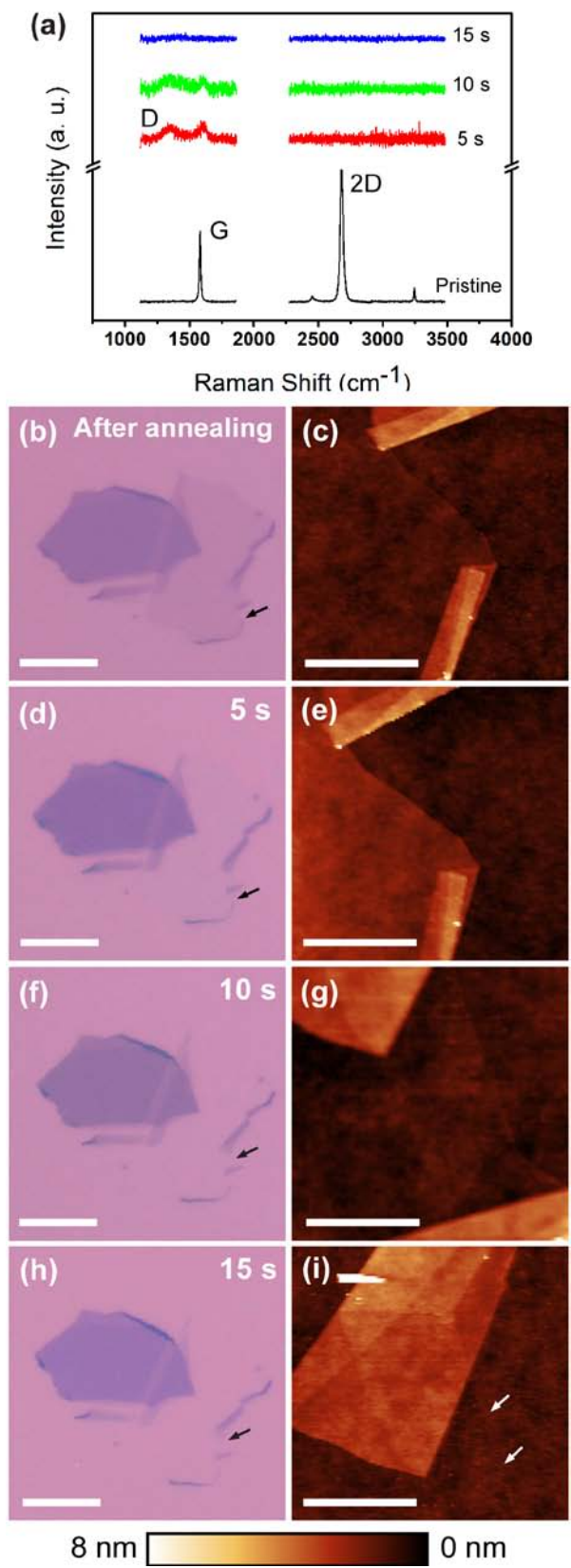

FIG. 3. Ar RIE etching of graphene. (a) Raman data showing the effect of each 5 sec etching pulse. Left column (b, d, f, h) are optical microscopy images of the graphene flake following

successive 5 s etching pulses (scale bar: 10 µm). Right column (c, e, g, i) are AFM images of the graphene flake, taken in region indicated by the black arrow in the correspondent optical image (scale bar: 1 µm). In (i), the white arrows indicate the original location of the monolayer edge.

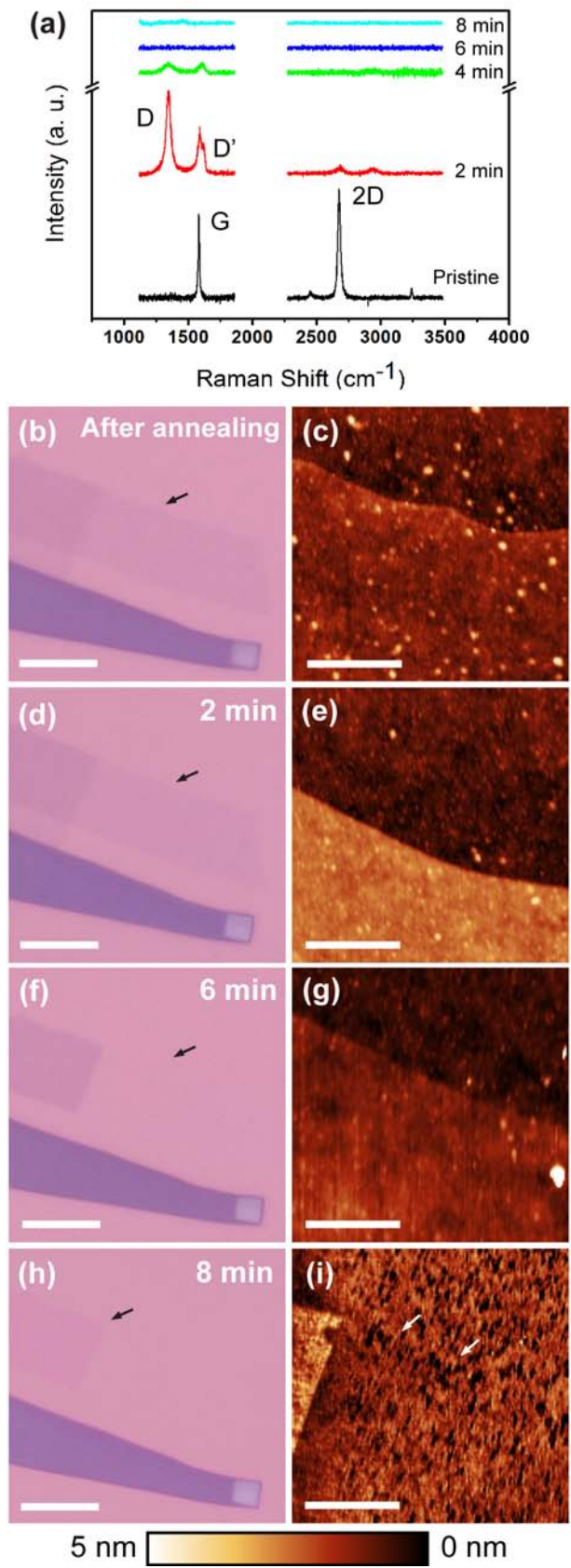

FIG. 4. Ar plasma etching of graphene. (a) Raman data showing the effect of each 2 min etching pulse. Left column (b, d, f, h) are optical microscopy images of the graphene flake

following successive 2 min etching steps (scale bar: 20 µm). Right column (c, e, g, i) are AFM images of the graphene flake, taken in the region indicated by the black arrow in the corresponding optical image (scale bar: 1 µm). In (i), the bilayer that was originally attached to the monolayer appears in the left side of the image as a reference, while the white arrows indicate the original location of the monolayer edge.

# Supplemental Material

# Optimization of graphene dry etching conditions via combined microscopic and spectroscopic analysis

Mariana C. Prado,[1,a)] Deep Jariwala,[2] Tobin J. Marks,[2,3] and Mark C. Hersam[2,3,b)]

[1]Departamento de Física, Universidade Federal de Minas Gerais, Av. Antônio Carlos, 6627, 31270-901 – Belo Horizonte, Brazil.

[2]Department of Materials Science and Engineering, Northwestern University, Evanston, Illinois, 60208, USA.

[3]Department of Chemistry, Northwestern University, Evanston, Illinois, 60208, USA.

[a)] This research was performed while M. Prado was at the Department of Materials Science and Engineering, Northwestern University, Evanston, Illinois, 60208, USA.

[b)] Author to whom correspondence should be addressed. Electronic mail: m-hersam@northwestern.edu

## 1 – Detailed Experimental Methods

Silicon with a 300 nm thick thermally grown $SiO_2$ layer was used as the substrate for graphene exfoliation. The Si wafers (Si orientation <100>) were purchased from Silicon Quest International. The wafers were doped n-type with As (resistivity = 0.001-0.005 $\Omega$-cm). Substrates were cleaned with sonication in acetone and isopropanol prior to deposition. Natural graphite from Nacional de Grafite was exfoliated using the scotch tape method.[1] Samples were mapped using an Olympus BX-51M optical microscope. Monolayers were identified by their optical contrast[2] and then confirmed by Raman spectroscopy.[3] Subsequently, the samples were annealed in a Lindberg blue tube furnace inside a quartz tube with ~1200 sccm flow of $H_2$/Ar 20% mixture at 300ºC for 3 h. A 1 hr ramp time was used to heat up and cool down the samples. This processing is necessary to remove glue residues from the exfoliation process, thereby facilitating the acquisition of AFM images (by preserving the tip from contamination) and avoiding interference from these residues in the etching processes.

Optical microscopy images and Raman spectra were reacquired after the annealing. The differences seen in the Raman signal before and after can be accounted for by the doping[4] induced by the annealing treatment. Since there is no change in the intensity of the defect-related peaks after the annealing process (i.e., there are no D or D' bands), the annealing process does not introduce any defects or damage in the graphene as seen in SM-Figure 1.

Atomic force microscopy (AFM) images were acquired with a CP Research (Thermomicroscopes) AFM operating in ambient conditions. All images were acquired in intermittent contact mode using standard commercial silicon cantilevers from Budget Sensors (All-in-One probes). The scanner allowed for a maximum 5 µm scan size. Images were analyzed with Gwyddion.[5]

Raman spectra were acquired using an Acton TriVista CRS Confocal Raman System from Princeton Instruments. The Ar 514.5 nm laser line was used with power < 2 mW to prevent sample heating. The 100x objective was used to focus the laser beam, and the Raman signal was dispersed on an 1800 gr/mm grating. The acquisition time was 10 sec for each spectrum for the pristine and post-annealing samples and 60 sec for the etched samples. Since etching decreases the Raman signal intensity, the increase in acquisition time was necessary to resolve the spectra and ensure that there was no more signal at a given point.

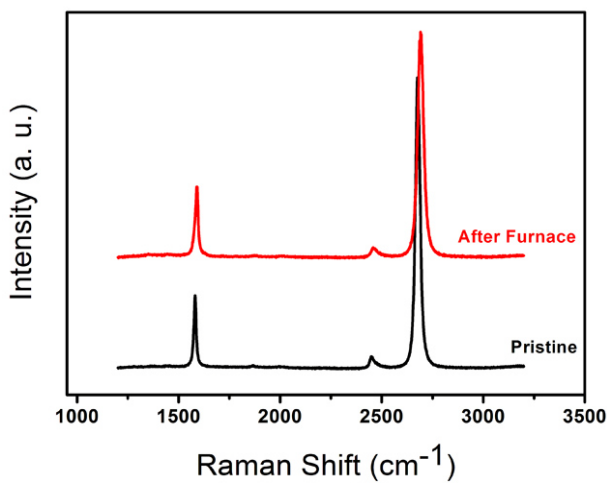

**SM-FIG. 1.** Raman spectrum before and after furnace annealing treatment to remove tape residues from a representative graphene sample.

Plasma etching was performed in a Plasma Cleaner PDC-001 from Harrick Plasma. Ultra-high purity $O_2$ or Ar were used as the working gases. The chamber was cleaned with an $O_2$ plasma for 15 min before each etching step. A power of 29.6 W was applied to the RF coil to generate the plasma. The gas pressure in the chamber was kept at 200 mTorr. Treatment time steps and pressure were optimized for each gas on earlier batches of samples.

Reactive ion etching (RIE) was performed using a Samco RIE-10 NR system. Ultra-high purity $O_2$ and Ar were used for etching. The chamber was pre-cleaned with a $CF_4 + O_2$ (50:50) mixture for 1 min at 100 W, followed by $O_2$ for 2 min at 50 W, and finally by Ar for 1 min at 50 W. This chamber cleaning procedure was repeated before each etching step on the samples. The graphene samples were etched under a gas flow rate of 10 sccm while maintaining a pressure of 26.5 Pa with a power of 30 W. In addition, the system applies ~50 V of DC bias during the etching process to accelerate the ions towards the samples.

For sample 1 ($O_2$ RIE), approximately the same region is shown in all four AFM images. This procedure required a great deal of time since the flakes are large (usually > 20 µm), and the maximum scan size is 5 µm. For all flakes, more than 1 region was scanned in each step, and no type of anisotropy was detected. After establishing the homogeneity of the etching, we imaged the edge of the flake in regions nearby but not necessarily the exact same location as the other figures. In addition, two or more monolayers per sample were characterized to verify reproducibility.

**2 – Ar Plasma Etching for 4 Minutes**

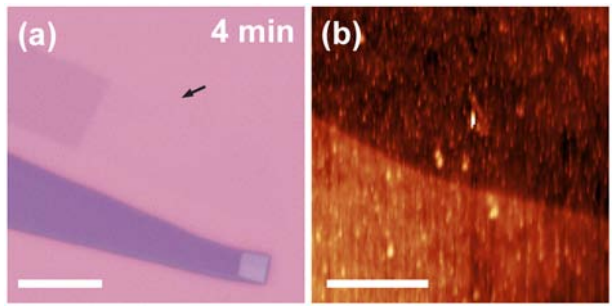

**SM-FIG. 2.** Supplement to Fig. 4. (a) Optical microscopy image and (b) AFM image of the graphene flake after a total of 4 min exposure to Ar plasma etching.

**3 - References**

[1] K. S. Novoselov, A. K. Geim, S. V. Morozov, D. Jiang, M. I. Katsnelson, I. V. Grigorieva, S. V. Dubonos, and A. A. Firsov, Nature **438**, 197 (2005).

[2] P. Blake, E. W. Hill, A. H. C. Neto, K. S. Novoselov, D. Jiang, R. Yang, T. J. Booth, and A. K. Geim, Appl. Phys. Lett. **91**, 063124 (2007).

[3] A. C. Ferrari, J. C. Meyer, V. Scardaci, C. Casiraghi, M. Lazzeri, F. Mauri, S. Piscanec, D. Jiang, K. S. Novoselov, S. Roth, and A. K. Geim, Phys. Rev. Lett. **97**, 187401 (2006).

[4] A. Das, S. Pisana, B. Chakraborty, S. Piscanec, S. K. Saha, U. V. Waghmare, K. S. Novoselov, H. R. Krishnamurthy, A. K. Geim, A. C. Ferrari, and A. K. Sood, Nat. Nanotechnol. **3**, 210 (2008).

[5] D. Necas and P. Klapetek, Cent. Eur. J. Phys. **10**, 181 (2012).